\documentclass[num-refs]{wiley-article}

\usepackage{siunitx}
\usepackage{multirow} 
\usepackage{booktabs}
\usepackage{tikz}
\usepackage{xcolor, colortbl}
\usepackage{pgfplots}
\pgfplotsset{compat=1.17}
\usepackage{algorithm}
\usepackage{algpseudocode}
\usepackage{adjustbox}

\title{Order of Addition in Mixture-Amount Experiments}

\abbrevs{OofA, Order-of-Addition; FDS,  Fraction of Design Space; PWo, Pair-Wise Ordering.}

\author[1\authfn{1}]{Taha Hasan}
\author[2\authfn{1}\authfn{2}]{Touqeer Ahmad}

\contrib[\authfn{1}]{Equally contributing authors.}

\affil[1]{Department of Statistics, Islamabad Model College for Boys, F-10/4, Islamabad, Pakistan.}
\affil[2]{ENSAI, CREST-UMR 9194, Bruz, France.}

\corraddress{\authfn{2}Touqeer Ahmad, ENSAI, CREST-UMR 9194, Bruz, France.}
\corremail{touqeer.ahmad@ensai.fr}

\fundinginfo{This study is not receiving any specific funding. }

\runningauthor{Hasan \& Ahmad.}

\newcommand{\answer}[1]{#1}
\newcommand{\answerR}[1]{#1}

\newcommand{\answerRR}[1]{#1}
\newcommand{\answerRrR}[1]{#1}

\begin{document}

\begin{frontmatter}
\maketitle

\begin{abstract}
In a mixture experiment, we study the behavior and properties of $m$ mixture components, where the primary focus is on the proportions of the components that make up the mixture rather than the total amount. Mixture-amount experiments are specialized types of mixture experiments where both the proportions of the components in the mixture and the total amount of the mixture are of interest. In this paper, we consider an Order-of-Addition (OofA) mixture-amount experiment in which the response depends on both the mixture amounts of components and their order of addition. Full mixture OofA designs are constructed to maintain orthogonality between the mixture-amount model terms and the effects of the order of addition. \answer{But the number of runs in such full OofA designs increases as $m$ increases. We employ the Threshold Accepting (TA) Algorithm to select an n-row subset from the full Order-of-Addition (OofA) mixture design that maximizes G-optimality while minimizing the number of experimental runs. Further, the G-efficiency criterion is used to assess how well the design supports the precise and unbiased estimation of the model parameters.} These designs enable the estimation of mixture-component model parameters and the order-of-addition effects.  The Fraction of Design Space (FDS) plot is used to provide a visual assessment of the prediction capabilities of a design across the entire design space.

\keywords{Mixture experiments, Order-of-addition, Mixture-Amount Experiment, G-
efficiency, Simplex-Lattice and Simplex-Centroid designs}
\end{abstract}
\end{frontmatter}

\section{Introduction }
\answer{The pharmaceutical industry relies heavily on research and development to produce effective and safe medications. Reducing costs and development time is crucial for maintaining competitiveness in this highly regulated and innovation-driven sector \citep{paul2010improve}. Medicines, which are carefully formulated mixtures of active pharmaceutical ingredients and excipients, must meet stringent quality standards to ensure therapeutic efficacy, stability, and patient safety. The optimization of these formulations requires a deep understanding of both the physicochemical properties of the components and their interactions.}

\answer{A critical challenge in pharmaceutical research is determining the optimal proportions of mixture components to achieve desired therapeutic effects while minimizing adverse reactions. This problem is analogous to other industries where mixture optimization is essential, such as in baking (where ingredient ratios affect texture and taste) or construction (where material blends determine strength and durability) \citep{cornell2002experiments}. Even minor variations in component proportions in pharmaceuticals can significantly alter a drug’s dissolution rate, bioavailability, and overall effectiveness \cite{montgomery2017design}.
}

\answer{ Mixture experiments provide a structured approach to optimizing formulations by systematically exploring how different component ratios influence key response variables, such as drug release kinetics or stability \citep{scheffe1958experiments}. An experiment in which the response variable $y$ depends on the proportion $x_i$ of components constituting a mixture blend, which is called a mixture experiment with the constraints $$0 \leq x_i \leq 1 \hspace{0.2cm} \text{and} \hspace{0.2cm} 
 \sum_{i=1}^m x_i; i=1, \dots, m.$$
The factor space of the mixture experiment in $m$ components is a $(m-1)$ dimensional simplex $S^{m-1}$. Factor spaces resulting from specified constraints on mixture components belong to the simplex $S^{m-1}$. The $n$ points in the simplex $S^{m-1}$ or its sub-region generate a mixture design with $n$ runs.
These $n$ runs are used to fit the mixture model to the observed responses. The design and analysis of such experiments enable researchers to identify the most effective formulations with minimal experimental runs, thereby reducing costs and development time \citep{myers2016response}.
}

\answer{A further extension of this approach is the mixture-amount experiment, where the component proportions and the total quantity of the mixture are varied to study their combined effects on the response. This is particularly relevant in drug development, where different dosages may be required for varying patient populations. For instance, the impact of changing the mixture component proportions and the total amount of the mixture on the response is measured by fitting a mixture-amount model to the design.} This study focuses on the order in which the mixture proportions \( x_1, x_2, \dots, x_m \) or each component amount \( a_1, a_2, \dots, a_m \) are added in each blend. We assume that the blends individually add the \( m \) components. Consequently, there are \( m! \) possible sequences of these proportions or amounts to investigate. This scenario is referred to as the order-of-addition (OofA) problem in a mixture-amount experiment. The response variable $y$ depends on the order in which the component's proportions or amounts are added in a composition and the interaction effect between the order and component proportions or amounts. In general, there are many scientific applications where the order of addition of the component proportions or amounts in a mixture blend affects the response. For instance, physical phenomena observed in science and engineering are influenced by the order in which $m$ components or materials are added. The first famous OofA experiment was conducted by \cite{fisher1971design}, which involved a lady who claimed to be able to differentiate whether tea or milk was added first to her cup. The experiment consisted of four replications of each tea order $\rightarrow$ milk and milk $\rightarrow$ tea. On the other hand, OofA experiments have also found wide application in biochemistry \citep{shinohara1998stimulation}, nutritional science \citep{karim2000effects}, bioengineering \citep{chandrasekaran2006substrate}, chemical experiments \citep{jiang2014sequential}, combinatorial drug therapy \citep{ding2015optimized}, and many others.

Limited literature exists on OofA mixture experiments. \citet{rios2022order} explored OofA mixture experiments and derived designs that account for interactions between the order of addition and mixture proportions. \answer{If a mixture design with $m$ components has $t$ runs, then the OofA mixture design constructed would have $t \times m!$ runs. The number of runs increases exponentially as $m$ increases. \citet{rios2022ta} used the Threshold Accepting (TA) algorithm to find a subset that has $t \times m!$ runs based on the D-optimality criterion.} However, no prior research has addressed OofA mixture-amount experiments and their design construction. \answer{In our study, we construct designs where both the order-of-addition and the total mixture amount are varied. Such OofA mixture component-amount designs would have $t$ x $m!$ mixture or component- order combinations. This paper focuses on designing cost-effective mixture-amount OofA experiments when funding constraints limit the number of runs. It is achieved by using the TA algorithm to select a limited set of runs from $t$ x $m!$ mixture or component-order combinations based on the G-optimality criterion. } These designs enable us to observe response changes for specific permutations of addition orders and total mixture amounts. The primary objective is determining the optimal proportions of ingredients to achieve the desired response, such as tablet weight and efficacy in formulating a pain relief tablet composed of active pharmaceutical ingredients, binders, and disintegrants. This involves designing experiments using factorial or mixture designs, identifying key variables like component proportions and total mixture amount, and measuring responses through statistical analysis to uncover significant effects on outcomes. The insights gained from these experiments can enhance product formulation and optimization, leading to improved performance and effectiveness across diverse applications, from developing optimal drug formulations to creating balanced nutrient profiles in fertilizers and food products.
\subsection{Models for Mixture-Amount Experiments}
The design for fitting the mixture-amount model is called the mixture-amount design, developed by \cite{piepel1985models}. The linear and quadratic expected response mixture-amount models for $m$ components are
\begin{equation*}
    E(y)=\sum_{i=1}^m \gamma_{i}^0 x_i + \sum_{i=1}^m \gamma_{i}^1 x_i A 
\end{equation*}
\begin{eqnarray}\label{mix_am_quad} 
   E(y)&=&  \sum_{i=1}^m \gamma_{i}^0 x_i+ \sum_{i<j}^m \gamma_{ij}^0 x_i x_j + \sum_{l=1}^2\left(\sum_{i=1}^m \gamma_{i}^l x_i+ \sum_{i<j}^m \gamma_{ij}^l x_i x_j\right)A^l.
\end{eqnarray}
This model consists of three second-order Scheff\`e polynomial forms in $m$ components, each multiplied by the powers of the total amount $A$ ($A^{0}=1$, $A$ and $A^{2}$). When $A = 0$, \answerR{this model predicts a zero response.} \citet{piepel1988note} modified this model to accommodate zero-amount condition. The alternative model uses the actual amounts of the components denoted by $a_1, a_2, \dots, a_m$ such that $a_1+a_2 \dots + a_m = A$. \answerR{The proportions $x_i$ are related to the amounts $a_i$ through 
$a_i = x_i A$.} This is called the component-amount model. The linear and quadratic expected response component-amount models are
\begin{eqnarray}
    E(y)&=& \alpha_0 +\sum_{i=1}^m \alpha_{i} a_i \\
    E(y)&=& \alpha_0 +\sum_{i=1}^m \alpha_{i} a_i + \sum_{i=1}^m \left(\alpha_{ii} a_i^2 + \sum_{i<j}^m \alpha_{ij} a_i a_j\right).
\end{eqnarray}
 For further discussion on mixture-amount and component-amount models, see, for instance,  \citet[pp. 403-418]{cornell2002experiments} 
 \subsection{Designs for Mixture-Amount Experiments}
 This subsection discusses the designs used to fit mixture-amount models formulated by expressing the parameters of a Scheffé-type mixture model as functions of a total amount variable. The standard designs employed for fitting these models include the simplex-lattice designs; however, other mixture designs can also be utilized. 
For instance, a $m$-ingredient simplex-lattice design, denoted by the notation $\{m, w\}$, consists of all valid mixture combinations which can be made for $m$ ingredients and with $w$ degree of the lattice, making the levels $0,1/w, 2/w,\dots, (w-1)/w$ and $w/w=1$. In general, it has $\binom{m+w-1}{w}$ design points.


Another mixture design, as discussed in \cite{cornell2002experiments}, is the simplex-centroid design. For an $m$-ingredient mixture, the simplex-centroid design consists of all vertices of the $(m - 1)$-dimensional simplex, $S^{m-1}$, and centroids formed by taking these vertices in combinations, such as pairs. This design includes $2^m - 1$ design points, starting with $m$ permutations of $(1, 0, \dots, 0)$, permutations of $\left(\frac{1}{2}, \frac{1}{2}, \dots, 0\right)$, permutations of $\left(\frac{1}{3}, \frac{1}{3}, \frac{1}{3}, \dots, 0\right)$, continuing until the centroid $\left(\frac{1}{m}, \frac{1}{m}, \dots, \frac{1}{m}\right)$ is reached. \citet{prescott2004mixture, prescott2008d} developed several component-amount designs by projecting standard symmetric mixture designs, achieved by removing one or more columns from the original design. Additionally, orthogonally blocked mixture designs based on Latin squares and F-squares blocking schemes were projected to create orthogonal blocked component-amount designs.


Any component-amount design with $m$ components can be derived from a simplex-lattice, simplex-centroid, or other standard mixture designs with $m + r$ ingredients by removing $r$ columns from the base design. For example, in a $\{4, 3\}$ simplex-lattice design, deleting one column ($r = 1$) produces a component-amount design with three ingredients and four levels of total amount $A$ (0, $1/3$, $2/3$, 1), denoted $\{4, 3\}_1$. Similarly, we can derive a component-amount design with projection to three dimensions (deleting any column) from a simplex-centroid design with $q = 4$ ingredients. The design has five levels of total amount $A$ ($A$ = 0, 1/2, 2/3, 3/4, and 1). \answerR{The design is given in Table~\ref{tab:Table 1}. For further discussion on the construction of component-amount designs via projections, see \citet{prescott2004mixture}.}
\begin{table}[htbp]
\centering
\captionsetup{justification=centerlast} 
\caption{Simplex-centroid design with four ingredients projected to a \\ component–amount design with three dimensions and five levels of $A$}
\scriptsize
\renewcommand{\arraystretch}{1.2}
\begin{tabular}{p{0.4cm} p{0.4cm} p{0.4cm} p{0.4cm} p{0.4cm} | p{0.4cm} p{0.4cm} p{0.4cm} p{0.4cm} p{0.4cm}}
\hline
\multicolumn{5}{|c|}{\parbox[c]{3cm}{\centering {Simplex–centroid design with \\ the proportions of 4 ingredients}}} &
\multicolumn{5}{c|}{\parbox[c]{3cm}{\centering {Component–amount design with\\ the amounts of 3 ingredients}}} \\
\hline
Run & $x_1$ & $x_2$ & $x_3$ & $x_4$ &
Run & $a_1$ & $a_2$ & $a_3$ & $A$ \\
\hline
1  & 1   & 0   & 0   & 0   & 1  & 1   & 0   & 0   & 1   \\
2  & 0   & 1   & 0   & 0   & 2  & 0   & 1   & 0   & 1   \\
3  & 0   & 0   & 1   & 0   & 3  & 0   & 0   & 1   & 1   \\
4  & 0   & 0   & 0   & 1   & 4  & 0   & 0   & 0   & 0   \\
5  & 1/2 & 1/2 & 0   & 0   & 5  & 1/2 & 1/2 & 0   & 1   \\
6  & 1/2 & 0   & 0   & 1/2 & 6  & 1/2 & 0   & 0   & 1   \\
7  & 1/2 & 0   & 1/2 & 0   & 7  & 1/2 & 0   & 1/2 & 1   \\
8  & 0   & 1/2 & 1/2 & 0   & 8  & 0   & 1/2 & 1/2 & 1   \\
9  & 0   & 1/2 & 0   & 1/2 & 9  & 0   & 1/2 & 0   & 1   \\
10 & 0   & 0   & 1/2 & 1/2 & 10 & 0   & 0   & 1/2 & 1   \\
11 & 1/3 & 1/3 & 1/3 & 0   & 11 & 1/3 & 1/3 & 1/3 & 1   \\
12 & 1/3 & 1/3 & 0   & 1/3 & 12 & 1/3 & 1/3 & 0   & 2/3 \\
13 & 1/3 & 0   & 1/3 & 1/3 & 13 & 1/3 & 0   & 1/3 & 2/3 \\
14 & 0   & 1/3 & 1/3 & 1/3 & 14 & 0   & 1/3 & 1/3 & 2/3 \\
15 & 1/4 & 1/4 & 1/4 & 1/4 & 15 & 1/4 & 1/4 & 1/4 & 3/4 \\
\hline
\end{tabular}
\label{tab:Table 1}
\end{table}

\indent The paper is structured as follows: Section \ref{OofA_Experiments} covers the OofA mixture experiment, the OofA mixture-amount experiment, and the related mixture amount and component-amount models. Section \ref{design_const} contains the construction of designs for mixture-amount and component-amount experiments, section \ref{application_designs} uses two examples for the application of designs, and section \ref{conclusion} concludes the paper.

\section{OofA Experiments}\label{OofA_Experiments}
The current literature on OofA experiments focuses primarily on the Pair-Wise Ordering model, first introduced by \citep{van1995design}. The Pair-Wise Ordering (PWO) model was formally named by \citet{voelkel2019design}. 
\answerR{Let $\mathbf{c} = (c_1, c_2, \dots, c_m)^T$ \answerRrR{denote} a permutation
of the $m$ components}.
Let \( \mathcal{P} \) be the set of all pairs \( (j, k) \) where \( 1 \leq j < k \leq m \).The PWO factor \( z_{jk}(\mathbf{c}) \) is defined as  
\begin{equation*}
    z_{jk}(\mathbf{c})=
    \begin{cases}
        1 & \text{if $j$ precedes $k$ in $c$}\\
        -1 & \text{if $k$ precedes $j$ in $c$},
    \end{cases}
\end{equation*}
hence if $\mathbf{c}=(2, 1, 3)$ then \( z_{12}(\mathbf{c})=-1 \), \( z_{13}(\mathbf{c})=1\) and \( z_{23}(\mathbf{c})=1 \).
The PWO model for the expected response $\eta(\mathbf{c})$ is given as
$$E(\mathbf{c})=\beta_0 +\sum_{jk\in S}z_{jk}(\mathbf{c}) \beta_{jk}.$$
The parameter $\beta_{jk}$ shows how the pair-wise order of components $j$ and $k$ affect the response. For finding the optimal order \citet{lin2019order} discussed the topological sorting methods for the PWO model. Several research studies can be seen in the literature on the Optimality of PWO experimental designs. See, for instance, \cite{peng2019design, winker2020construction}.
\subsection{The OofA mixture-amount experiments}
\label{the_OofA_mixture-amount_experiments}
An OofA mixture-amount experiment is a type of mixture experiment where we consider not only the effects of the component amounts on the response but also the effects of the order in which the components are added. The response surface would be a function of component amounts $a_1, a_2,\dots,a_m$ and the order of their addition in a mixture. We test the significant effect of the order of addition of components and the amounts of components on the response. Our concern in this study is how to incorporate OofA effects into the mixture-amount and component-amount models. Therefore, we construct the OofA mixture component-amount designs using mixture simplex designs. We use the modified PWO notations given by \citet{rios2022order}. Define $(m-1)$ dimensional simplex $S$ and $\mathcal{P}$ be the set of all pairs $(j, k)$ where $1 \leq j <k \leq m$. To get the full design matrix for fitting the mixture-amount model, we define PWO factor $z_{jk} (x, \mathbf{c})$ as
\begin{equation*}
	 z_{jk}(x,\mathbf{c}) =
	\begin{cases} 
		1 &x_j, x_i \neq 0 \hspace{0.1cm}  \text{$j$ is before $k$ in $c$}\\
  0 & x_j = 0  \hspace{0.1cm} or \hspace{0.1cm}x_i=0\\
  - 1 & x_j, x_i \neq 0  \hspace{0.1cm} \text{$k$ is before $j$ in $c$}.\\
	\end{cases}
\end{equation*}
Now, to get a complete design matrix for fitting the component-amount model, the PWO factor $z_{jk} (a, c)$ for a permutation $c$ of $(1,2,\dots, m)$ components amounts is defined as
\begin{equation*}
	 z_{jk}(a,\mathbf{c}) =
	\begin{cases} 
		1 &a_j, a_i \neq 0 \hspace{0.1cm}  \text{$j$ is before $k$ in $c$}\\
  0 & a_j = 0  \hspace{0.1cm} or \hspace{0.1cm}a_k=0\\
  - 1 & a_j, a_i \neq 0  \hspace{0.1cm} \text{$k$ is before $j$ in $c$}\\
	\end{cases}
\end{equation*}
\subsection{Models for OofA mixture-amount experiments}
We further constructed the model for the PWO and mixture component amounts. Without interaction between the mixture amount and PWO factor, the expected response additive model is developed as \begin{eqnarray}\label{mix_lin}
    E(y)&=&\sum_{i=1}^m \gamma_{i}^0 x_i + \sum_{k<l} \delta_{kl}^0 Z_{kl} + \sum_{i=1}^m \gamma_{i}^1 x_i A+ \sum_{k<l} \delta_{kl}^1 Z_{kl}A.
\end{eqnarray}
After including mixture-order interaction, the expected response model becomes
\begin{eqnarray}\label{mix_quad}
    E(y)&=&\sum_{i=1}^m \gamma_{i}^0 x_i + 
     \sum_{i<j} \gamma_{ij}^0 x_i x_j +
    \sum_{k<l} \delta_{kl}^0 Z_{kl} +\sum_{i} \sum_{k<l \hspace{0.1cm}\text{or}\hspace{0.1cm} k>l, i=k} \lambda_{i(kl)}^t x_i Z_{kl}  \nonumber \\ 
    &+& 
    \sum_{t=1}^2\left(\sum_{i=1}^m \gamma_i^t x_i +\sum_{i<j}^m \gamma_{ij}^t x_i x_j +\sum_{k<l}^m \delta_{kl}^t Z_{kl}+ \sum_{i} \sum_{k<l \hspace{0.1cm}\text{or} \hspace{0.1cm}k>l, i=k} \lambda_{i(kl)}^t x_i Z_{kl} \right)A^t.
\end{eqnarray}
On the other hand, the expected response additive model for the component amount and PWO factor without interaction is  
\begin{eqnarray}\label{comp_lin}
    E(y)&=&\alpha_0 +\sum_{i=1}^m \alpha_i a_{i} + \sum_{k<l} \delta_{kl} Z_{kl}.
\end{eqnarray}
After including component-amount and order interaction, the model has the following form
\begin{eqnarray}\label{order_inter}\label{com_quad}
    E(y)&=&\alpha_0 +\sum_{i=1}^m \alpha_i a_{i} + \sum_{k<l} \beta_{kl} Z_{kl} +\sum_{i=1}^m \alpha_{ii} a_{i}^2 + \sum_{i<j} \gamma_{ij} a_i a_j \nonumber \\
    &+& \sum_{i} \sum_{k<l \hspace{0.1cm}\text{or} \hspace{0.1cm}k>l, k\neq l } \delta_{i(kl)} a_i Z_{kl}.
\end{eqnarray}
\subsection{TA algorithm for OofA mixture-amount designs}
\citet{dueck1990threshold} proposed a heuristic optimization algorithm called the TA algorithm. The modified algorithm for this study is given in Algorithm~\ref {alg:TA_Goptimal}. 
It is widely applied to solving experimental design problems.
This algorithm allows worse solutions to be accepted if only they fall within the threshold. Over time, the threshold is reduced, leading the algorithm toward a more refined solution (see Algorithm~\ref{alg:threshold}).  \citet{rios2022ta} used TA algorithm to determine D-optimal OofA mixture designs. The D-optimality criterion maximizes the determinant of the information matrix. We use the TA algorithm to determine G-optimal OofA mixture-amount design with optimal number of runs \( n \geq p \), where $p$ is the number of model parameters.
The G-optimality criterion minimizes the maximum prediction variance over the design space.
\[
\text{G-optimality =} \quad \min_{\xi} \left( \max_{x \in \mathcal{X}} \mathbf{f}^\top(x) \left[ \mathbf{F}^\top \mathbf{F} \right]^{-1} \mathbf{f}(x) \right)
\]
\noindent
where \( \xi \) denotes the experimental design, \( \mathcal{X} \) is the design space, \( \mathbf{f}(x) \) is the model vector evaluated at point \( x \), and \( \mathbf{F} \) represents the model matrix constructed from the selected design points.

\begin{algorithm}
\caption{Empirical Threshold Sequence Generation}
\label{alg:threshold}
\begin{algorithmic}[1]
\State \textbf{Input:} Number of iterations $n_{\text{iter}}$, full design $D_{\text{full}}$, design size $n$
\State \textbf{Output:} Empirical threshold sequence $\mathcal{T}^*$
\vspace{0.2cm}
\State Initialize $D_0 \subset D_{\text{full}}$ such that $|D_0| = n$
\State Initialize empty set $\mathcal{T} \gets \emptyset$
\For{$r = 1$ to $n_{\text{iter}}$}
    \State \answerRrR{Generate the design $D_1$ by randomly exchanging $k$ rows in $D_0$ with $k$ rows from $D_{\text{full}} \setminus D_{0}$.} 
    \State Compute efficiency difference as $
    t_r = \left| \text{Efficiency}(D_0) - \text{Efficiency}(D_1) \right|$
    \State Update threshold set: $\mathcal{T} \gets \mathcal{T} \cup \{t_r\}$
\EndFor
\State Sort $\mathcal{T}$ in descending order: $\mathcal{T}_{\text{sorted}}$
\State Keep top 50\% values $
\mathcal{T}^* = \left\{ t_1, t_2, \dots, t_{\lfloor n_{\text{iter}}/2 \rfloor} \right\} $
\State \textbf{Return} $\mathcal{T}^*$
\end{algorithmic}
\end{algorithm}

\begin{algorithm}
\caption{TA Algorithm for Finding G-Optimal OofA Mixture-Amount Design}
\label{alg:TA_Goptimal}
\begin{algorithmic}[1]
\State \textbf{Input:} Full design $D_{\text{full}}$, number of design points $n$, number of iterations $n_{\text{iter}}$, model function $f(x)$, threshold sequence $\{t_r\}$
\State \textbf{Output:} Final design $D^*$ with improved G-efficiency
\vspace{0.2cm}

\State Randomly select $n$ rows from $D_{\text{full}}$ to form the initial design $D_0$
\State Construct the model matrix $M_0$ for $D_0$
\State Compute the inverse information matrix as
$
C_0 = (M_0^T M_0)^{-1}
$
\State Compute the initial maximum prediction variance as
$
v_{\max}(D_0) = \max_{x \in D_{\text{full}}} f(x)^T C_0 f(x)
$
\For{$r = 1$ to $n_{\text{iter}}$}
    \State Initialize: $D_1 \gets D_0$, $C_1 \gets C_0$, $v_{\max}(D_1) \gets v_{\max}(D_0)$
    \State Generate a neighbor design $D_1 \in N_k(D_0)$ by randomly exchanging $k$ rows in $D_0$
    
    \For{each exchange $(x \in D_0, y \in D_{\text{full}} \setminus D_0)$}
        \State Compute $v(y) = f(y)^T C_0 f(y)$
        \State Update inverse matrix using Sherman–Morrison formula:
        \[
        C_1 = C_0 + C_0 F_1 (I_2 + F_2^T C_0 F_1)^{-1} F_2^T C_0
        \]
        \State Compute:
        $
        v_{\max}(D_1) = \max_{x \in D_{\text{full}}} f(x)^T C_1 f(x)
        $
    \EndFor

    \If{$v_{\max}(D_1) < v_{\max}(D_0) + t_r$}
        \State Accept $D_1$ as the new current design: $D_0 \gets D_1$, $C_0 \gets C_1$
    \EndIf
\EndFor

\State \textbf{Return:} Final design $D^* = D_0$ with associated G-efficiency
\end{algorithmic}
\end{algorithm}

\section{Formulation of new designs}\label{design_const}

In this section, we first construct the OofA component-amount experiment. Then, we construct the designs that are suitable for fitting mixture-amount and component-amount models with PWO variables.
\subsection{Construction of OofA mixture-amount design}
We construct OofA mixture component-amount designs using mixture simplex designs. The modified PWO notations are used as given in section \ref{the_OofA_mixture-amount_experiments}. The \{3, 3\} simplex-lattice design is used as the base design. We further defined three PWO variables $z_{12}$, $z_{13}$, and $z_{23}$, and as a result, the design matrix has 21 design points. It can be used for fitting the models given in \eqref{mix_lin} and \eqref{mix_quad}. The constructed design matrix is given in Table~\ref{tab:Table 2}.
\begin{table}[htbp]
    \centering
    \captionsetup{justification=centerlast} 
    \caption{The OofA simplex-lattice design with $m = 3$ and $l = 3$.}
    \scriptsize
    \begin{tabular}{ p{1cm} p{1cm} p{1cm} p{1cm} p{1cm} p{1cm} p{1cm}}
        \hline
        Runs & $x_1$ & $x_2$ & $x_3$ & $z_{12}$ & $z_{13}$ & $z_{23}$ \\
        \hline
  1 & 1 & 0 & 0 & 0 & 0 & 0 \\
  2 & 0 & 1 & 0 & 0 & 0 & 0 \\
  3 & 0 & 0 & 1 & 0 & 0 & 0 \\
  4 & 0.33 & 0.67 & 0 & 1 & 0 & 0 \\
  5 & 0.33 & 0.67 & 0 & -1 & 0 & 0 \\
  6 & 0.33 & 0 & 0.67 & 0 & 1 & 0 \\
  7 & 0.33 & 0 & 0.67 & 0 & -1 & 0 \\
  8 & 0.67 & 0.33 & 0 & -1 & 0 & 0 \\
  9 & 0.67 & 0.33 & 0 & 1 & 0 & 0 \\
  10 & 0.67 & 0 & 0.33 & 0 & -1 & 0 \\
  11 & 0.67 & 0 & 0.33 & 0 & 1 & 0 \\
  12 & 0 & 0.33 & 0.67 & 0 & 0 & 1 \\
  13 & 0 & 0.33 & 0.67 & 0 & 0 & -1 \\
  14 & 0 & 0.67 & 0.33 & 0 & 0 & -1 \\
  15 & 0 & 0.67 & 0.33 & 0 & 0 & 1 \\
  16 & 0.33 & 0.33 & 0.33 & 1 & 1 & 1 \\
  17 & 0.33 & 0.33 & 0.33 & 1 & 1 & -1 \\
  18 & 0.33 & 0.33 & 0.33 & 1 & -1 & -1 \\
  19 & 0.33 & 0.33 & 0.33 & -1 & 1 & 1 \\
  20 & 0.33 & 0.33 & 0.33 & -1 & -1 & 1 \\
  21 & 0.33 & 0.33 & 0.33 & -1 & -1 & -1 \\
  \hline
\end{tabular}\label{tab:Table 2}
\end{table}

\subsection{Construction of OofA component-amount design}
A three-instrument component-amount design with five levels of the total amount $(A = 0, 1/2, 2/3, 3/4, 1)$ is developed by projecting a simplex centroid design with four ingredients. This is done \answerR{by following the methods used in \cite{prescott2004mixture}.} Using this as the base design, we derived a three-component  OofA component-amount design with 31 design points and included three PWO variables $z_{12}$, $z_{13}$, and $z_{23}$.The constructed design is given in Table~\ref{tab:Table 3}. In addition, we can fit the models \eqref{comp_lin} and \eqref{com_quad}, and we can analyze the design.  
\begin{table}[htbp]
   \centering
   \captionsetup{width=0.75\linewidth}
    \caption{
The OofA mixture component-amount design for three ingredients and five levels of total amount \(A\).
}
\scriptsize
\centering
\begin{tabular}{ p{1cm} p{1cm} p{1cm} p{1cm} p{1cm} p{1cm} p{1cm} p{1cm} p{1cm} p{1cm} }
  \hline
 Runs & $a_1$ & $a_2$ & $a_3$ & $z_{12}$ & $z_{13}$ & $z_{23}$ & $A$ \\
  \hline
  1  & 1    & 0    & 0    & 0    & 0    & 0    & 1   \\
  2  & 0    & 1    & 0    & 0    & 0    & 0    & 1   \\
  3  & 0    & 0    & 1    & 0    & 0    & 0    & 1   \\
  4  & 1/2  & 1/2  & 0    & 1    & 0    & 0    & 1   \\
  5  & 1/2  & 1/2  & 0    & -1   & 0    & 0    & 1   \\
  6  & 1/2  & 0    & 1/2  & 0    & 1    & 0    & 1   \\
  7  & 1/2  & 0    & 1/2  & 0    & -1   & 0    & 1   \\
  8  & 0    & 1/2  & 1/2  & 0    & 0    & 1    & 1   \\
  9  & 0    & 1/2  & 1/2  & 0    & 0    & -1   & 1   \\
  10 & 1/3  & 1/3  & 1/3  & 1    & 1    & 1    & 1   \\
  11 & 1/3  & 1/3  & 1/3  & 1    & 1    & -1   & 1   \\
  12 & 1/3  & 1/3  & 1/3  & 1    & -1   & -1   & 1   \\
  13 & 1/3  & 1/3  & 1/3  & -1   & 1    & 1    & 1   \\
  14 & 1/3  & 1/3  & 1/3  & -1   & -1   & 1    & 1   \\
  15 & 1/3  & 1/3  & 1/3  & -1   & -1   & -1   & 1   \\
  16 & 1/4  & 1/4  & 1/4  & 1    & 1    & 1    & 3/4 \\
  17 & 1/4  & 1/4  & 1/4  & 1    & 1    & -1   & 3/4 \\
  18 & 1/4  & 1/4  & 1/4  & 1    & -1   & -1   & 3/4 \\
  19 & 1/4  & 1/4  & 1/4  & -1   & 1    & 1    & 3/4 \\
  20 & 1/4  & 1/4  & 1/4  & -1   & -1   & 1    & 3/4 \\
  21 & 1/4  & 1/4  & 1/4  & -1   & -1   & -1   & 3/4 \\
  22 & 1/3  & 1/3  & 0    & 1    & 0    & 0    & 2/3 \\
  23 & 1/3  & 1/3  & 0    & -1   & 0    & 0    & 2/3 \\
  24 & 1/3  & 0    & 1/3  & 0    & 1    & 0    & 2/3 \\
  25 & 1/3  & 0    & 1/3  & 0    & -1   & 0    & 2/3 \\
  26 & 0    & 1/3  & 1/3  & 0    & 0    & 1    & 2/3 \\
  27 & 0    & 1/3  & 1/3  & 0    & 0    & -1   & 2/3 \\
  28 & 1/2  & 0    & 0    & 0    & 0    & 0    & 1/2 \\
  29 & 0    & 1/2  & 0    & 0    & 0    & 0    & 1/2 \\
  30 & 0    & 0    & 1/2  & 0    & 0    & 0    & 1/2 \\
  31 & 0    & 0    & 0    & 0    & 0    & 0    & 0   \\
  \hline
\end{tabular}\label{tab:Table 3}
\end{table}
\answerR{A general construction of OofA mixture-amount and component amount design is given in the Algorithm~\ref{alg:oofa_design}.}

\begin{algorithm}
\caption{General Construction of OofA Mixture-Amount and Component-Amount Design Matrices}
\label{alg:oofa_design}
\begin{algorithmic}[1]
\State \textbf{Input:} Number of ingredients \( m \), total amount set \( A \), initial design matrix \( D \).
\State \textbf{Output:} Final design $D^*$ with improved G-efficiency.
\State \answerRrR{Let $\mathcal{C}$ be the set of all $m!$ permutations of ($c_1, \dots , c_m)$.}
\State For each \( 1 \leq j < k \leq m \), define pairwise order variables:
\[
z_{jk}(x,\mathbf{c}) =
	\begin{cases} 
		1 &x_j, x_i \neq 0 \hspace{0.1cm}  \text{$j$ is before $k$ in $c$}\\
  0 & x_j = 0  \hspace{0.1cm} or \hspace{0.1cm}x_i=0\\
  - 1 & x_j, x_i \neq 0  \hspace{0.1cm} \text{$k$ is before $j$ in $c$}.\\
	\end{cases}
    \text{and} \quad
     z_{jk}(a,\mathbf{c}) =
	\begin{cases} 
		1 &a_j, a_i \neq 0 \hspace{0.1cm}  \text{$j$ is before $k$ in $c$}\\
  0 & a_j = 0  \hspace{0.1cm} or \hspace{0.1cm}a_k=0\\
  - 1 & a_j, a_i \neq 0  \hspace{0.1cm} \text{$k$ is before $j$ in $c$}\\
	\end{cases}
\]
\State Construct extended design space:
\[
\mathcal{D}_{\text{OofA}} = \left\{ (\mathbf{x}, a, \mathbf{c}, \{z_{jk}\}) : \mathbf{x} \in \Delta^{m-1}, a \in A, \mathbf{c} \in \mathcal{C} \right\}
\]
\State Apply TA-algorithm \ref{alg:TA_Goptimal} to find G-optimal design with reduced runs.
\State \textbf{Output:} Optimized OofA design matrix \( D^* \).
\end{algorithmic}
\end{algorithm}

\section{Application of OofA mixture-amount designs}\label{application_designs}
The OofA mixture-amount designs developed in Section~\ref{design_const} have been implemented in practical, real-world scenarios. These designs have been applied to real-life examples to demonstrate their effectiveness in capturing both the influence of component amounts and the sequence in which the components are added to the final response. Two real-life examples are considered in the following sections. 
\subsection{Study of the joint action of three related hormones in groups of mice}
This example is taken from \citet{cornell2002experiments}. The presented results of an experiment involved administering $10$ different blends of three distinct hormones to $10$ groups of $12$ mice each. The detail can be found in \citet{claringbold1955use}. The hormone blends were also prepared in three amounts: $0.75 \times 10^{-4} \mu g$, $1.50 \times 10^{-4}\mu g$, and $3.00 \times 10{-4}\mu g$, resulting in a total of $30$ groups. The primary focus is to determine how different mixtures of hormones influenced the cornification response of the vaginal epithelium in ovariectomized mice. \answerR{The response ($y$) referred to the percent response, which represents the proportion of mice (out of 12) in each group that responded to hormone injection. It was measured for each combination of mixtures at different levels of total hormone amount $A$ $(0.75, 1.5$ and $3.0)$. The $10$ hormone blends at each amount $A$ form a $\{3, 3\}$ simplex lattice, creating a $\{3, 3\}$ x $3$ mixture amount design. A quadratic mixture-amount model ~\eqref{mix_am_quad} was fitted and it was observed that the effect on the percent response of increasing the amount of hormone $3$ is less than with hormone $1$ and hormone $2$.}  Now we consider the order of addition in the experiment. So the OofA $\{3, 3\}$ simplex-lattice design now has $21$ mixture blends ($x_1, x_2, x_3$,$z_{12}, z_{13}$ and $z_{23}$) including PWO variables at each total amount $A$.This creates a $63$ blend OofA mixture-amount design given in Table~\ref{tab:Table 4}. The objective of the analysis includes how the proportions and order of addition impacted the response when we fit the mixture-amount model~\eqref{mix_quad}. For model identifiability, the mixture-order interactions $x_1 z_{13}, x_2z_{12}, x_3z_{23}$ are ignored.
\begin{table}[!htbp]
\centering
\caption{The OofA simplex-lattice design with \( m = 3 \) and \( l = 3 \) with three levels of total amount \( A \)}
\scriptsize
\centering
\begin{tabular}{p{0.5cm} p{0.5cm} p{0.5cm} p{0.5cm} p{0.5cm} p{0.5cm} p{0.5cm} p{0.5cm} | p{0.5cm} p{0.5cm} p{0.5cm} p{0.5cm} p{0.5cm} p{0.5cm} p{0.5cm} p{0.5cm}}
  \hline
 Runs & $x_1$ & $x_2$ & $x_3$ & $z_{12}$ & $z_{13}$ & $z_{23}$ & $A$ & Runs & $x_1$ & $x_2$ & $x_3$ & $z_{12}$ & $z_{13}$ & $z_{23}$ & $A$ \\
  \hline
1  & 1    & 0    & 0    & 0    & 0    & 0    & 0.75 & 33 & 0    & 0.33 & 0.67 & 0    & 0    & 1    & 1.50 \\
2  & 0    & 1    & 0    & 0    & 0    & 0    & 0.75 & 34 & 0    & 0.33 & 0.67 & 0    & 0    & -1   & 1.50 \\
3  & 0    & 0    & 1    & 0    & 0    & 0    & 0.75 & 35 & 0    & 0.67 & 0.33 & 0    & 0    & -1   & 1.50 \\
4  & 0.33 & 0.67 & 0    & 1    & 0    & 0    & 0.75 & 36 & 0    & 0.67 & 0.33 & 0    & 0    & 1    & 1.50 \\
5  & 0.33 & 0.67 & 0    & -1   & 0    & 0    & 0.75 & 37 & 0.33 & 0.33 & 0.33 & 1    & 1    & 1    & 1.50 \\
6  & 0.67 & 0.33 & 0    & -1   & 0    & 0    & 0.75 & 38 & 0.33 & 0.33 & 0.33 & 1    & 1    & -1   & 1.50 \\
7  & 0.67 & 0.33 & 0    & 1    & 0    & 0    & 0.75 & 39 & 0.33 & 0.33 & 0.33 & 1    & -1   & -1   & 1.50 \\
8  & 0.33 & 0    & 0.67 & 0    & 1    & 0    & 0.75 & 40 & 0.33 & 0.33 & 0.33 & -1   & 1    & 1    & 1.50 \\
9  & 0.33 & 0    & 0.67 & 0    & -1   & 0    & 0.75 & 41 & 0.33 & 0.33 & 0.33 & -1   & -1   & 1    & 1.50 \\
10 & 0.67 & 0    & 0.33 & 0    & -1   & 0    & 0.75 & 42 & 0.33 & 0.33 & 0.33 & -1   & -1   & -1   & 1.50 \\
11 & 0.67 & 0    & 0.33 & 0    & 1    & 0    & 0.75 & 43 & 1    & 0    & 0    & 0    & 0    & 0    & 3.00 \\
12 & 0    & 0.33 & 0.67 & 0    & 0    & 1    & 0.75 & 44 & 0    & 1    & 0    & 0    & 0    & 0    & 3.00 \\
13 & 0    & 0.33 & 0.67 & 0    & 0    & -1   & 0.75 & 45 & 0    & 0    & 1    & 0    & 0    & 0    & 3.00 \\
14 & 0    & 0.67 & 0.33 & 0    & 0    & -1   & 0.75 & 46 & 0.33 & 0.67 & 0    & 1    & 0    & 0    & 3.00 \\
15 & 0    & 0.67 & 0.33 & 0    & 0    & 1    & 0.75 & 47 & 0.33 & 0.67 & 0    & -1   & 0    & 0    & 3.00 \\
16 & 0.33 & 0.33 & 0.33 & 1    & 1    & 1    & 0.75 & 48 & 0.67 & 0.33 & 0    & -1   & 0    & 0    & 3.00 \\
17 & 0.33 & 0.33 & 0.33 & 1    & 1    & -1   & 0.75 & 49 & 0.67 & 0.33 & 0    & 1    & 0    & 0    & 3.00 \\
18 & 0.33 & 0.33 & 0.33 & 1    & -1   & -1   & 0.75 & 50 & 0.33 & 0    & 0.67 & 0    & 1    & 0    & 3.00 \\
19 & 0.33 & 0.33 & 0.33 & -1   & 1    & 1    & 0.75 & 51 & 0.33 & 0    & 0.67 & 0    & -1   & 0    & 3.00 \\
20 & 0.33 & 0.33 & 0.33 & -1   & -1   & 1    & 0.75 & 52 & 0.67 & 0    & 0.33 & 0    & -1   & 0    & 3.00 \\
21 & 0.33 & 0.33 & 0.33 & -1   & -1   & -1   & 0.75 & 53 & 0.67 & 0    & 0.33 & 0    & 1    & 0    & 3.00 \\
22 & 1    & 0    & 0    & 0    & 0    & 0    & 1.50 & 54 & 0    & 0.33 & 0.67 & 0    & 0    & 1    & 3.00 \\
23 & 0    & 1    & 0    & 0    & 0    & 0    & 1.50 & 55 & 0    & 0.33 & 0.67 & 0    & 0    & -1   & 3.00 \\
24 & 0    & 0    & 1    & 0    & 0    & 0    & 1.50 & 56 & 0    & 0.67 & 0.33 & 0    & 0    & -1   & 3.00 \\
25 & 0.33 & 0.67 & 0    & 1    & 0    & 0    & 1.50 & 57 & 0    & 0.67 & 0.33 & 0    & 0    & 1    & 3.00 \\
26 & 0.33 & 0.67 & 0    & -1   & 0    & 0    & 1.50 & 58 & 0.33 & 0.33 & 0.33 & 1    & 1    & 1    & 3.00 \\
27 & 0.67 & 0.33 & 0    & -1   & 0    & 0    & 1.50 & 59 & 0.33 & 0.33 & 0.33 & 1    & 1    & -1   & 3.00 \\
28 & 0.67 & 0.33 & 0    & 1    & 0    & 0    & 1.50 & 60 & 0.33 & 0.33 & 0.33 & 1    & -1   & -1   & 3.00 \\
29 & 0.33 & 0    & 0.67 & 0    & 1    & 0    & 1.50 & 61 & 0.33 & 0.33 & 0.33 & -1   & 1    & 1    & 3.00 \\
30 & 0.33 & 0    & 0.67 & 0    & -1   & 0    & 1.50 & 62 & 0.33 & 0.33 & 0.33 & -1   & -1   & 1    & 3.00 \\
31 & 0.67 & 0    & 0.33 & 0    & -1   & 0    & 1.50 & 63 & 0.33 & 0.33 & 0.33 & -1   & -1   & -1   & 3.00 \\
32 & 0.67 & 0    & 0.33 & 0    & 1    & 0    & 1.50 & --  & --   & --   & --   & --   & --   & --   & -- \\
\hline
\end{tabular}\label{tab:Table 4}
\end{table}
\\
\indent
\answer{Since the design has two many runs, we select a subset of runs from our full design. We implemented the TA algorithm in Python to generate a G-optimal OofA mixture-amount design using 10,000 iterations. The G-optimal OofA mixture amount design generated via the TA algorithm.} \answerR{The design has 36 runs, which is the minimum runs required to estimate the model parameters ~\eqref{mix_quad}. We simulated the percent responses ($y$) for each mixture based on the results of \citet{claringbold1955use, cornell2002experiments}.The design is given in Table~\ref{tab:Table 5}.  The simulated responses are consistent with the findings reported in \citet{claringbold1955use, cornell2002experiments}. We fit the OofA mixture-amount model~\eqref{mix_quad} to the design given in Table~\ref{tab:Table 5}. We used ANOVA to assess the significance of the effects of mixture proportions with variable total amounts and their order-of-addition effect on the response. The results are presented in Table ~\ref{tab:Table 6}. At the 0.05 significance level, it can be observed that both the hormone proportions at different levels of total amount and their order of addition have a significant effect on the percentage response (proportion of mice out of 12). 
 The analysis of the design reveals that the maximum prediction variance of the design was found to be 1.00, while the average prediction variance of the design was 0.89.  A G-efficiency of 89\% implies that the design performs well in terms of minimizing the worst-case prediction variance relative to the average prediction variance. The design provides reliable and stable predictions throughout the mixture space, making it a strong candidate to accurately model response surfaces.}

\begin{table}[htbp]
\centering
\captionsetup{width=0.75\linewidth}
\caption{G-optimal OofA mixture component–amount design obtained via TA algorithm}
\scriptsize
\centering
\renewcommand{\arraystretch}{1.2}
\setlength{\tabcolsep}{3pt} 

\begin{tabular}{p{0.4cm} p{0.4cm} p{0.4cm} p{0.4cm} p{0.4cm} p{0.4cm} p{0.4cm} p{0.4cm} p{0.4cm} | p{0.4cm} p{0.4cm} p{0.4cm} p{0.4cm} p{0.4cm} p{0.4cm} p{0.4cm} p{0.4cm} p{0.4cm}}
\toprule
Run & $x_1$ & $x_2$ & $x_3$ & $z_{12}$ & $z_{13}$ & $z_{23}$ & $A$ & $y$ & Run & $x_1$ & $x_2$ & $x_3$ & $z_{12}$ & $z_{13}$ & $z_{23}$ & $A$ & $y$ \\
\midrule
1  & 0 & 0.67 & 0.33 & 0  & 0  & 1  & 3.0  & 56 & 19 & 1 & 0 & 0 & 0 & 0 & 0 & 0.75 & 40 \\
2  & 0 & 0 & 1 & 0 & 0 & 0 & 1.5 & 52 & 20 & 0.67 & 0.33 & 0 & -1 & 0 & 0 & 3.0 & 15 \\
3  & 0.67 & 0.33 & 0 & 1 & 0 & 0 & 0.75 & 29 & 21 & 0.67 & 0 & 0.33 & 0 & 1 & 0 & 3.0 & 54 \\
4  & 0.67 & 0.33 & 0 & -1 & 0 & 0 & 0.75 & 95 & 22 & 0 & 0.33 & 0.67 & 0 & 0 & 1 & 3.0 & 19 \\
5  & 1 & 0 & 0 & 0 & 0 & 0 & 3.0 & 19 & 23 & 0 & 1 & 0 & 0 & 0 & 0 & 0.75 & 59 \\
6  & 0.33 & 0 & 0.67 & 0 & 1 & 0 & 0.75 & 38 & 24 & 0.33 & 0.33 & 0.33 & 1 & -1 & -1 & 3.0 & 25 \\
7  & 0.33 & 0 & 0.67 & -1 & 0 & 0 & 0.75 & 26 & 25 & 0.67 & 0.33 & 0 & 1 & 0 & 0 & 0.75 & 94 \\
8  & 0 & 1 & 0 & 0 & 0 & 0 & 0.75 & 50 & 26 & 0.33 & 0.33 & 0.33 & 1 & 1 & 1 & 0.75 & 42 \\
9  & 0 & 0.33 & 0.67 & 0 & 0 & 1 & 0.75 & 50 & 27 & 0.33 & 0.33 & 0.33 & -1 & 1 & 1 & 3.0 & 76 \\
10 & 0.33 & 0 & 0.67 & 0 & -1 & 0 & 3.0 & 63 & 28 & 0 & 1 & 0 & 0 & 0 & 0 & 1.5 & 34 \\
11 & 0 & 1 & 0 & 0 & 0 & 0 & 0.75 & 95 & 29 & 0.33 & 0.33 & 0.33 & 1 & 1 & 1 & 3.0 & 45 \\
12 & 0.33 & 0 & 0.67 & 0 & 1 & 0 & 1.5 & 45 & 30 & 0 & 1 & 0 & 0 & 0 & 0 & 3.0 & 43 \\
13 & 0.67 & 0.33 & 0 & -1 & 0 & 0 & 1.5 & 33 & 31 & 0.33 & 0.33 & 0.33 & -1 & -1 & 1 & 3.0 & 60 \\
14 & 0 & 0.33 & 0.67 & 0 & 0 & -1 & 0.75 & 22 & 32 & 0.33 & 0.33 & 0.33 & 1 & -1 & -1 & 3.0 & 43 \\
15 & 0 & 0.33 & 0.67 & 0 & 0 & -1 & 1.5 & 51 & 33 & 0.67 & 0 & 0.33 & 0 & 1 & 0 & 1.5 & 42 \\
16 & 0.67 & 0.33 & 0 & 1 & 0 & 0 & 1.5 & 39 & 34 & 0.33 & 0.33 & 0.33 & -1 & 1 & 1 & 1.5 & 35 \\
17 & 0.33 & 0.33 & 0.33 & -1 & -1 & -1 & 1.5 & 36 & 35 & 0.33 & 0 & 0.67 & 0 & 1 & 0 & 3.0 & 40 \\
18 & 0 & 0 & 1 & 0 & 0 & 0 & 1.5 & 73 & 36 & 0.33 & 1 & 0 & 0 & 0 & 0 & 3.0 & 9 \\
\bottomrule
\end{tabular}
\label{tab:Table 5}
\end{table}
\begin{table}[htbp]
\centering
\captionsetup{width=0.75\linewidth}
\caption{Analysis and parameter estimates of OofA mixture–amount model}
\scriptsize
\centering
\renewcommand{\arraystretch}{1.2}
\begin{tabular}{p{1cm} p{1cm} p{1cm} p{1cm} | p{1cm} p{1cm} p{1cm} p{1cm}}
\hline
\textbf{Term} & \textbf{Estimate} & \textbf{Std.Err} & \textbf{t value} &
\textbf{Term} & \textbf{Estimate} & \textbf{Std.Err} & \textbf{t value} \\
\hline
$x_1$ & -73.28 & 10.56 & -6.940 & $z_{13}A$ & -6.32 & 43.84 & -0.144 \\
$x_2$ & -45.92 & 17.15 & -2.667 & $z_{23}A$ & -30.19 & 43.54 & -0.693 \\
$x_3$ & -35.78 & 13.06 & -2.737 & $x_1x_2A$ & -4.11 & 26.82 & -0.153 \\
$z_{12}$ & 31.22 & 14.52 & 2.388 & $x_1x_3A$ & 194.43 & 171.28 & 1.135 \\
$z_{13}$ & -7.13 & 24.19 & -0.294 & $x_2x_3A$ & 56.77 & 44.10  & 1.287 \\
$z_{23}$ & -24.14 & 11.98 & -2.015 & $x_1z_{12}A$ & -5.68 & 42.15 & -0.135 \\
$x_1x_2$ & -45.21 & 37.21 & -1.215 & $x_2z_{23}A$ & 21.31 & 20.28 & 1.050 \\
$x_1x_3$ & 47.77 & 30.05 & 2.589 & $x_1A^2$ & -15.73 & 12.64 & -1.224 \\
$x_2x_3$ & -45.66 & 57.60 & -0.792 & $x_2A^2$ & -48.82 & 83.10 & -0.527 \\
$x_1z_{12}$ & -25.71 & 11.41 & -2.253 & $x_3A^2$ & 773.37 & 719.76 & 1.070 \\
$x_3z_{13}$ & -8.96 & 69.19 & -0.129 & $z_{12}A^2$ & 151.87 & 56.28 & 2.690 \\
$x_2z_{23}$ & -23.05 & 14.18 & -1.625 & $z_{13}A^2$ & -5.59 & 42.37 & -0.132 \\
$x_1A$ & -14.85 & 11.79 & -1.259 & $z_{23}A^2$ & -33.46 & 41.37 & -0.808 \\
$x_2A$ & 53.47 & 28.64 & 1.866 & $x_1x_2A^2$ & -19.85 & 62.96 & -0.315 \\
$x_3A$ & 753.39 & 768.98 & 0.979 & $x_1x_3A^2$ & -221.50 & 177.09 & -1.250 \\
$z_{12}A$ & 128.99 & 64.24 & 2.008 & $x_2x_3A^2$ & 68.41 & 77.64 & 0.881 \\
\hline
\end{tabular}
\label{tab:Table 6}
\end{table}

\begin{figure}[t]
    \centering
\includegraphics[width=0.5\linewidth]{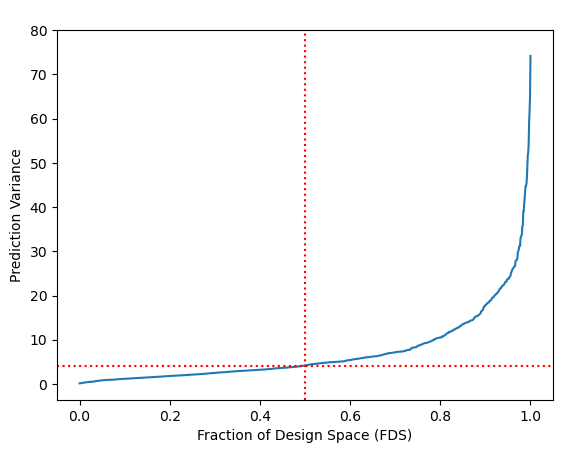}
    \caption{\centering FDS Plot for OofA mixture-amount model}
    \label{fig:mix-amount}
\end{figure}

For visual assessment of the prediction capability of the design, we use the fraction of the design space plot (FDS), as shown in Figure~\ref{fig:mix-amount}. The FDS plot indicates that the design is robust, with a large portion of design space having low prediction variance. This suggests reliable and consistent predictions for most factor combinations.
\answer{A simulation study using 10,000 synthetically generated datasets demonstrated that the reduced 36-run design, obtained by the Threshold Accepting (TA) algorithm, maintained predictive accuracy comparable to the full 63-run design. Although the G-efficiency of the full design was higher (4520.3)  due to redundancy, the TA design consistently exhibited strong mean G-efficiency (2451.3), with a substantially reduced experimental burden (36 design points). The mean normalized G-efficiency for TA Design is 51.3\% which means that the 36-point design still retains about 51\% of the prediction capability of the full design. This validates the practical utility of the TA-algorithm based design, showing that it can deliver robust performance with fewer experimental runs.} \\
\begin{figure}[t]
    \centering
\includegraphics[width=0.5\linewidth]{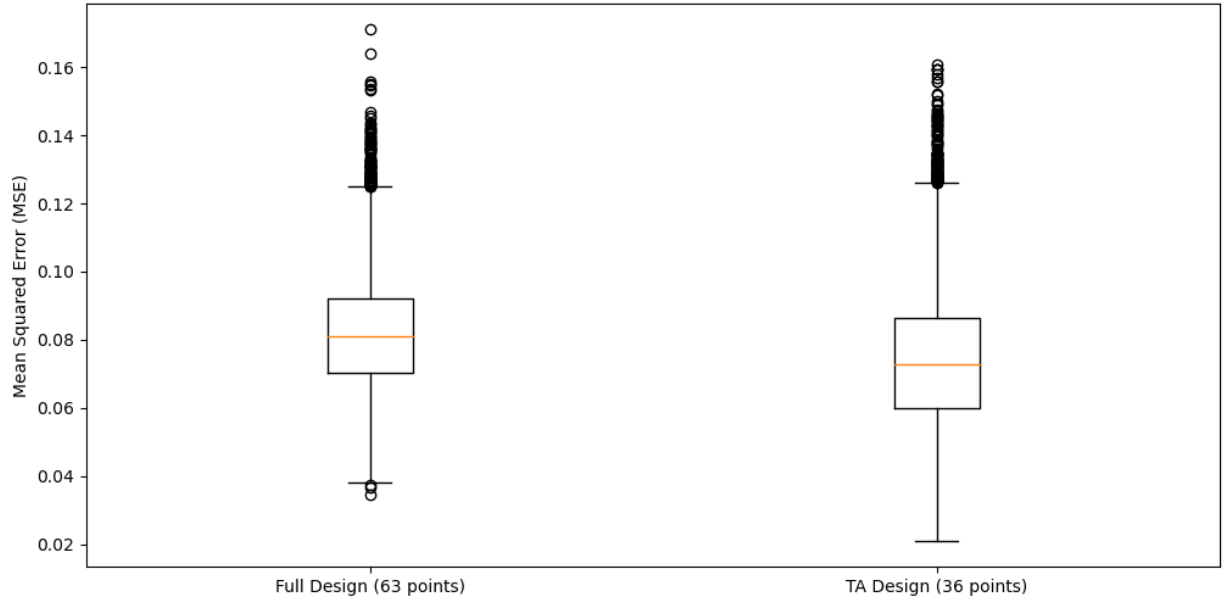}
    \caption{\Centering Boxplot of MSEs across 10,000 simulated datasets}
    \label{fig:Box_plot}
\end{figure}
\answerRrR{The boxplot of MSEs for 10,000 simulations in Figure ~\ref{fig:Box_plot}  shows that the TA-reduced design (36 points) achieves a lower median MSE compared to the full design (63 points), while exhibiting a slightly wider spread of errors. This indicates that the TA algorithm can efficiently select design points, maintaining predictive accuracy with substantially fewer runs.}
\subsection{ A Placebo tablet formulation with three components}
It may be easier to explain dosage units such as tablets or gelatin capsules by the amount of each component rather than their proportions when the total mass varies. However, using proportions and the variable total amount in the model is usually more convenient.
This example is taken from \citet{lewis1991pharmaceutical}. \citet{huisman1984development} have discussed the use of a mixture of experimental designs to prepare placebo tablets. Their formulation included up to three diluents: $\alpha$-lactose monohydrate, potato starch, and anhydrous $\alpha$-lactose. The tablets also had magnesium stearate as a lubricant, but since this was kept at a constant low level, the rest of the formulation, made up entirely of diluents, was considered 100\%. \answerR{The placebo tablets were prepared by introducing 500 mg of the mixture. Measured responses were crushing strength (N), disintegration time (s), and price (cts / kg). A simplex lattice design for a special cubic model with three checkpoints was used. In this study, we consider both the variable dosage strength and the order of addition of components, focusing on a single response, i.e., tablet crushing strength (N). The OofA component-amount design for a quadratic model ~\eqref{order_inter} with 31 design points is used. The design is given in Table ~\ref{tab:Table 7} with five levels of total amount i.e. 0, 250, 333.3, 375 and 500. The zero amount situation is used for the control test and is discussed in \citet{piepel1988note}.}
\begin{table}[t]
\centering
\caption{The OofA mixture component-amount design for three ingredients and five levels of total amount $A$.}
\scriptsize
\centering
\begin{tabular}{p{0.5cm} p{0.5cm} p{0.5cm} p{0.5cm} p{0.5cm} p{0.5cm} p{0.5cm} p{0.5cm} | p{0.5cm} p{0.5cm} p{0.5cm} p{0.5cm} p{0.5cm} p{0.5cm} p{0.5cm} p{0.5cm}}
  \hline
 Runs & $a_1$ & $a_2$ & $a_3$ & $z_{12}$ & $z_{13}$ & $z_{23}$ & $A$ & Runs & $a_1$ & $a_2$ & $a_3$ & $z_{12}$ & $z_{13}$ & $z_{23}$ & $A$ \\
  \hline
1  & 500    & 0    & 0    & 0 & 0 & 0 & 500 & 17 & 125   & 125   & 125   & 1 & 1 & -1 & 375 \\
2  & 0      & 500  & 0    & 0 & 0 & 0 & 500 & 18 & 125   & 125   & 125   & 1 & -1 & -1 & 375 \\
3  & 0      & 0    & 500  & 0 & 0 & 0 & 500 & 19 & 125   & 125   & 125   & -1 & 1 & 1 & 375 \\
4  & 250    & 250  & 0    & 1 & 0 & 0 & 500 & 20 & 125   & 125   & 125   & -1 & -1 & 1 & 375 \\
5  & 250    & 250  & 0    & -1 & 0 & 0 & 500 & 21 & 125   & 125   & 125   & -1 & -1 & -1 & 375 \\
6  & 250    & 0    & 250  & 0 & 1 & 0 & 500 & 22 & 166.7 & 166.7 & 0     & 1 & 0 & 0 & 333.3 \\
7  & 250    & 0    & 250  & 0 & -1 & 0 & 500 & 23 & 166.7 & 166.7 & 0     & -1 & 0 & 0 & 333.3 \\
8  & 0      & 250  & 250  & 0 & 0 & 1 & 500 & 24 & 166.7 & 0     & 166.7 & 0 & 1 & 0 & 333.3 \\
9  & 0      & 250  & 250  & 0 & 0 & -1 & 500 & 25 & 166.7 & 0     & 166.7 & 0 & -1 & 0 & 333.3 \\
10 & 166.7  & 166.7 & 166.7 & 1 & 1 & 1 & 500 & 26 & 0     & 166.7 & 166.7 & 0 & 0 & 1 & 333.3 \\
11 & 166.7  & 166.7 & 166.7 & 1 & 1 & -1 & 500 & 27 & 0     & 166.7 & 166.7 & 0 & 0 & -1 & 333.3 \\
12 & 166.7  & 166.7 & 166.7 & 1 & -1 & -1 & 500 & 28 & 250   & 0     & 0     & 0 & 0 & 0 & 250 \\
13 & 166.7  & 166.7 & 166.7 & -1 & 1 & 1 & 500 & 29 & 0     & 250   & 0     & 0 & 0 & 0 & 250 \\
14 & 166.7  & 166.7 & 166.7 & -1 & -1 & 1 & 500 & 30 & 0     & 0     & 250   & 0 & 0 & 0 & 250 \\
15 & 166.7  & 166.7 & 166.7 & -1 & -1 & -1 & 500 & 31 & 0     & 0     & 0     & 0 & 0 & 0 & 250 \\
16 & 125    & 125  & 125  & 1 & 1 & 1 & 375 & --- & ---   & ---   & ---   & --- & --- & --- & --- \\
\hline
\end{tabular}\label{tab:Table 7}
\end{table}
\\
\indent
\answer{Now we use the TA algorithm and construct a G-optimal design with 18 optimal number of runs. We simulate the response variable (y), i.e., the crushing strength (N), for all design points in the TA algorithm-based G-optimal design.  The design is given in Table~\ref{tab:Table 8}. The average of the variances of predicted responses across all design points is 1, indicating that the overall uncertainty in the predictions is uniformly distributed throughout the design space. Since we have 89\% G-efficiency, it implies that the design performs very well under the G-optimality criterion and ensures optimal precision even for the worst-case point in the design space.}  \answerR{We fit the model  ~\eqref{order_inter} to the design and use ANOVA to assess the significance of mixture amount and order effects on the response. The results are presented in Table ~\ref{tab:Table 9}. \answerRrR{Using $\alpha$ = 0.05, it can be observed} that component amounts and the order of addition have a significant effect on the crushing strength.
}

\begin{table}[htbp]
   \centering
\captionsetup{justification=centerlast} 
\caption{G-optimal Component-amount OofA design obtained via TA algorithm}
\scriptsize
\centering
\begin{tabular}{ p{1cm} p{1cm} p{1cm} p{1cm} p{1cm} p{1cm} p{1cm} p{1cm} p{1cm} }
  \hline
  Run & $a_1$ & $a_2$ & $a_3$ & $z_{12}$ & $z_{13}$ & $z_{23}$ & $A$ & $y$\\
  \hline
  1 & 0 & 0 & 0 & 0 & 0 & 0 & 0 & 0 \\
  2 & 250 & 250 & 0 & -1 & 0 & 0 & 500 & 63.5 \\
  3 & 250 & 250 & 0 & 1 & 0 & 0 & 500 & 120.4 \\
  4 & 0 & 500 & 0 & 0 & 0 & 0 & 500 & 98.5\\
  5 & 250 & 0 & 250 & 0 & -1 & 0 & 500 & 44.7 \\
  6 & 125 & 125 & 125 & -1 & 1 & 1 & 375 & 68.7\\
  7 & 500 & 0 & 0 & 0 & 0 & 0 & 500 & 77.2\\
  8 & 250 & 0 & 250 & 0 & 1 & 0 & 500 & 130.6\\
  9 & 0 & 250 & 0 & 0 & 0 & 0 & 250 & 46.3\\
  10 & 0 & 250 & 250 & 0 & 0 & 1 & 500 & 68.6\\
  11 & 0 & 0 & 250 & 0 & 0 & 0 & 250 & 50.4 \\
  12 & 125 & 125 & 125 & 1 & 1 & -1 & 375 & 61.3\\
  13 & 250 & 0 & 0 & 0 & 0 & 0 & 250 & 26.7\\
  14 & 0 & 250 & 250 & 0 & 0 & -1 & 500 & 49.0\\
  15 & 125 & 125 & 125 & -1 & -1 & -1 & 375 & 86.7\\
  16 & 0 & 0 & 500 & 0 & 0 & 0 & 500 & 78.7 \\
  17 & 125 & 125 & 125 & 1 & -1 & -1 & 375 & 80.4 \\
  18 & 166.7 & 0 & 166.7 & 0 & -1 & 0 & 333.3 & 39.4 \\
  \hline
\end{tabular}\label{tab:Table 8}
\end{table}
\begin{table}[htbp]
\centering
\captionsetup{width=0.75\linewidth}
\caption{Parameter estimates for the OofA quadratic component–amount model}
\scriptsize
\renewcommand{\arraystretch}{1.2}
\begin{tabular}{p{1cm} p{1cm} p{1cm} p{1cm} | p{1cm} p{1cm} p{1cm} p{1cm}}
\hline
\textbf{Term} & \textbf{Estimate} & \textbf{Std.Err} & \textbf{t value} &
\textbf{Term} & \textbf{Estimate} & \textbf{Std.Err} & \textbf{t value} \\
\hline
$a_0$     & 109.99 & 30.35 & 3.620 & $a_2^2$     & -4.20  & 15.77 & -0.266 \\
$a_1$     & 56.65  & 27.57 & 2.054 & $a_3^2$     & -14.99 & 14.66 & -1.022 \\
$a_2$     & 17.07  & 38.35 & 0.445 & $a_1a_2$    & 12.51  & 23.97 & 0.522 \\
$a_3$     & 61.02  & 29.57 & 2.063 & $a_1a_3$    & 5.54   & 23.41 & 0.236 \\
$z_{12}$  & 28.45  & 9.66  & 2.945 & $a_2a_3$    & -44.34 & 18.97 & -2.337 \\
$z_{13}$  & 44.02  & 9.50  & 4.633 & $a_1z_{12}$ & 75.34  & 25.92 & 2.901 \\
$z_{23}$  & 9.80   & 9.66  & 1.014 & $a_3z_{13}$ & 92.86  & 25.93 & 3.582 \\
$a_1^2$   & 7.96   & 14.96 & 0.532 & $a_2z_{23}$ & 42.78  & 28.68 & 1.491 \\
\hline
\end{tabular}
\label{tab:Table 9}
\end{table}

\answer{The FDS plot in Figure~\ref{fig:comp-amount} shows that the design performs well across most of the space, with relatively low prediction variance for a large portion. The higher prediction variances are limited to a smaller portion of the design space.\\
\begin{figure}[t]
    \centering
    \includegraphics[width=0.5\linewidth]{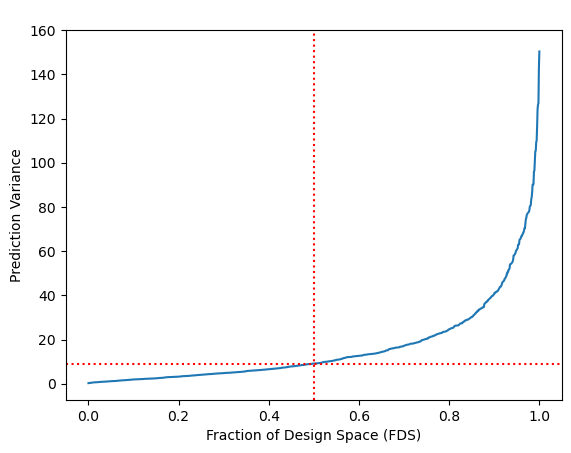}
    \caption{\centering FDS Plot for OofA component-amount model.}
    \label{fig:comp-amount}
\end{figure}
\indent In a simulation study generating 10,000 designs, each with 36 runs, the Threshold Accepting (TA) algorithm was used to produce 16-run reduced designs. The full design had a mean G-efficiency value 1485.0, while the TA-reduced design achieved a mean G-efficiency value 934. The mean normalized G-efficiency for the TA design was 63\%, indicating that the 16-run design retained approximately half of the predictive capacity of the full design. This demonstrates the TA algorithm's effectiveness in selecting efficient design points, providing a substantial reduction in experimental runs without compromising performance, thereby validating its practical utility in experimental design.}
\begin{figure}[t]
    \centering
\includegraphics[width=0.5\linewidth]{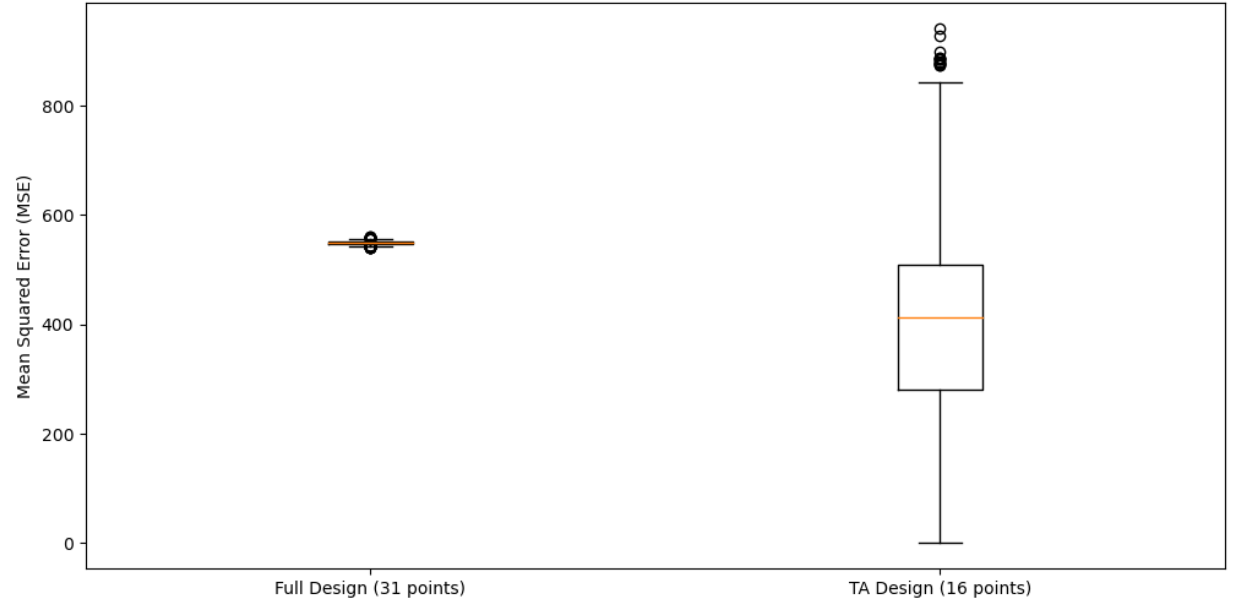}
    \caption{\Centering Boxplot of MSEs across 10,000 simulated datasets}
    \label{fig:Box_plot_comp}
\end{figure}
\answerRrR{The Boxplot in Figure~\ref{fig:Box_plot_comp} shows that the TA-reduced design (16 points) achieves a lower median MSE than the full design (31 points), but with a wider spread of errors across simulations. This shows that the TA algorithm can substantially reduce the number of runs while maintaining predictive accuracy, although with increased variability due to the smaller sample size}
\section{Conclusion}\label{conclusion}
In this study, we developed OofA mixture-amount experiments where the component amounts and the order of their addition to the mixture affect the overall response. We modified mixture-amount and component-amount models by incorporating the order-of-addition effect. To fit these models, we constructed designs for the OofA mixture-amount experiment.
\answer{But such OofA mixture-amount designs have a large number of runs. To make the designs practical and cost effective, we used the TA algorithm that uses a subset of $t$ x $m!$ runs, which results in a much higher G-efficiency relative to the full OofA mixture-amount design. This reduction in run size improves design efficiency compared to the full OofA mixture-amount design, which contains many replicates. As recommended by \cite{rios2022ta}, we have used the target run size $n \geq p$ , where $p$ is the number of model parameters. } 
\\  The designs constructed in this paper can be used to determine whether the order and the mixture amounts have a significant effect on the response. These full mixture-amount OofA designs confirmed orthogonality between mixture-amount model terms and OofA effects. This allows for a precise estimation of both the parameters of the mixture components and the effects of order of addition. \\
\indent The first example is the study of the joint action of three related hormones in groups of mice.  An OofA mixture-amount design, with three levels of total amount A, was constructed. As the design has a large number of runs, we used the TA algorithm and selected a subset of runs, which is a G-optimal design with high \answerRrR{G-efficiency value 89\%.} This indicates that the design is highly efficient and suitable for most practical applications.\\
\indent In the second example, we generated a three-factor OofA component amount design with 31 runs and with five levels of total amount.  We applied the TA algorithm and selected a partial fraction of the full design. It is a G-optimal design with the highest \answerRrR{G-efficiency value 89\%} and a minimum number of runs. The design achieves a good trade-off between efficiency and robustness.  
\\
\indent The developed OofA designs provide a systematic approach to include order-of-addition effects in mixture-amount experiments, enabling a more comprehensive design space exploration and better optimization of the response variable. It is the initial step for designing OofA mixture-amount experiments. \answerRR{We note that order-of-addition effects are most readily examined under controlled laboratory or pilot-
scale conditions, and in practice, significant findings from such studies can inform the development of
standardized procedures that facilitate consistent implementation during manufacturing.} \answer{This work can be extended for constrained and orthogonally blocked mixture-amount experiments. Many other optimality criteria can be used for research interest. Metaheuristic algorithms, which do not require candidate points, can be used to construct optimal designs.}

\section*{Acknowledgements}
We thank the anonymous reviewers for their comments that helped us to improve the quality of the paper. Touqeer Ahmad acknowledges support from the R\`egion Bretagne through Project SAD-2021- MaEVa.

\section*{Conflict of interest}
The authors declare no conflict of interest. 
\section*{Authors contribution}
Both authors equally contributed to this research. 
 \section*{Data availability statement }
The data and code used in this study are available from the corresponding author upon reasonable request.


\printendnotes

\bibliography{sample}

\begin{thebibliography}{26}
\providecommand{\natexlab}[1]{#1}
\providecommand{\url}[1]{\texttt{#1}}
\providecommand{\urlprefix}{}

\bibitem[{Paul et~al.(2010)Paul, Steven M and Mytelka, Daniel S and Dunwiddie, Christopher T and Persinger, Charles C and Munos, Bernard H and Lindborg, Stacy R and Schacht, Aaron L}]{paul2010improve}
Paul SM, Mytelka DS, Dunwiddie CT, Persinger CC, Munos BH, Lindborg SR, et~al.
\newblock How to improve R\&D productivity: the pharmaceutical industry's grand challenge.
\newblock Nature reviews Drug discovery 2010;9(3):203--214.

\bibitem[{Cornell(2002)Cornell, John A.}]{cornell2002experiments}
Cornell JA.
\newblock Experiments with Mixtures: Designs, Models, and the Analysis of Mixture Data.
\newblock 3rd ed. New York, USA: John Wiley \& Sons; 2002.

\bibitem[{Montgomery(2017)Montgomery, Douglas C}]{montgomery2017design}
Montgomery DC.
\newblock Design and analysis of experiments.
\newblock John wiley \& sons; 2017.

\bibitem[{Scheff{\'e}(1958)Scheff{\'e}, Henry}]{scheffe1958experiments}
Scheff{\'e} H.
\newblock Experiments with mixtures.
\newblock Journal of the Royal Statistical Society: Series B (Methodological) 1958;20(2):344--360.

\bibitem[{Myers et~al.(2016)Myers, Raymond H and Montgomery, Douglas C and Anderson-Cook, Christine M}]{myers2016response}
Myers RH, Montgomery DC, Anderson-Cook CM.
\newblock Response surface methodology: process and product optimization using designed experiments.
\newblock John Wiley \& Sons; 2016.

\bibitem[{Fisher(1971)Fisher, R. A.}]{fisher1971design}
Fisher RA.
\newblock The Design of Experiments.
\newblock 9th ed. London, UK: Macmillan; 1971.

\bibitem[{Shinohara and Ogawa(1998)Shinohara, Akira and Ogawa, Tomoko}]{shinohara1998stimulation}
Shinohara A, Ogawa T.
\newblock Stimulation by Rad52 of yeast Rad51-mediated recombination.
\newblock Nature 1998;391(6665):404--407.

\bibitem[{Karim et~al.(2000)Karim, Malina and McCormick, Kellie and Kappagoda, C Tissa}]{karim2000effects}
Karim M, McCormick K, Kappagoda CT.
\newblock Effects of cocoa extracts on endothelium-dependent relaxation.
\newblock The Journal of nutrition 2000;130(8):2105S--2108S.

\bibitem[{Chandrasekaran et~al.(2006)Chandrasekaran, Sangeetha M and Bhartiya, Sharad and Wangikar, Pramod P}]{chandrasekaran2006substrate}
Chandrasekaran SM, Bhartiya S, Wangikar PP.
\newblock Substrate specificity of lipases in alkoxycarbonylation reaction: QSAR model development and experimental validation.
\newblock Biotechnology and bioengineering 2006;94(3):554--564.

\bibitem[{Jiang and Ng(2014)Jiang, Xiong-Jie and Ng, Dennis KP}]{jiang2014sequential}
Jiang XJ, Ng DK.
\newblock Sequential logic operations with a molecular keypad lock with four inputs and dual fluorescence outputs.
\newblock Angewandte Chemie International Edition 2014;53(39):10481--10484.

\bibitem[{Ding et~al.(2015)Ding, Xianting and Matsuo, Kyle and Xu, Lin and Yang, Jian and Zheng, Longpo}]{ding2015optimized}
Ding X, Matsuo K, Xu L, Yang J, Zheng L.
\newblock Optimized combinations of bortezomib, camptothecin, and doxorubicin show increased efficacy and reduced toxicity in treating oral cancer.
\newblock Anti-cancer drugs 2015;26(5):547--554.

\bibitem[{Rios and Lin(2022)Rios, Nicholas and Lin, Dennis KJ}]{rios2022order}
Rios N, Lin DK.
\newblock Order-of-addition mixture experiments.
\newblock Journal of Quality Technology 2022;54(5):517--526.

\bibitem[{Rios et~al.(2022)Rios, Nicholas and Winker, Peter and Lin, Dennis KJ}]{rios2022ta}
Rios N, Winker P, Lin DK.
\newblock TA algorithms for D-optimal OofA Mixture designs.
\newblock Computational Statistics \& Data Analysis 2022;168:107411.

\bibitem[{Piepel and Cornell(1985)Piepel, Gregory F and Cornell, John A}]{piepel1985models}
Piepel GF, Cornell JA.
\newblock Models for mixture experiments when the response depends on the total amount.
\newblock Technometrics 1985;27(3):219--227.

\bibitem[{Piepel(1988)Piepel, Gregory F}]{piepel1988note}
Piepel GF.
\newblock A note on models for mixture-amount experiments when the total amount takes a zero value.
\newblock Technometrics 1988;30(4):449--450.

\bibitem[{Prescott and Draper(2004)Prescott, Philip and Draper, Norman R}]{prescott2004mixture}
Prescott P, Draper NR.
\newblock Mixture component-amount designs via projections, including orthogonally blocked designs.
\newblock Journal of Quality Technology 2004;36(4):413--431.

\bibitem[{Prescott and Draper(2008)Prescott, Philip and Draper, Norman R}]{prescott2008d}
Prescott P, Draper NR.
\newblock D-optimal mixture component-amount designs for quadratic and cubic models.
\newblock Journal of Applied Statistics 2008;35(7):739--749.

\bibitem[{Van~Nostrand(1995)Van Nostrand, RC}]{van1995design}
Van~Nostrand R.
\newblock Design of experiments where the order of addition is important.
\newblock In: ASA proceedings of the Section on Physical and Engineering Sciences, vol. 155 American Statistical Association Alexandria, VA, USA; 1995. p. 160.

\bibitem[{Voelkel(2019)Voelkel, Joseph G}]{voelkel2019design}
Voelkel JG.
\newblock The design of order-of-addition experiments.
\newblock Journal of Quality Technology 2019;51(3):230--241.

\bibitem[{Lin and Peng(2019)Lin, Dennis KJ and Peng, Jiayu}]{lin2019order}
Lin DK, Peng J.
\newblock Order-of-addition experiments: A review and some new thoughts.
\newblock Quality Engineering 2019;31(1):49--59.

\bibitem[{Peng et~al.(2019)Peng, Jiayu and Mukerjee, Rahul and Lin, Dennis KJ}]{peng2019design}
Peng J, Mukerjee R, Lin DK.
\newblock Design of order-of-addition experiments.
\newblock Biometrika 2019;106(3):683--694.

\bibitem[{Winker et~al.(2020)Winker, Peter and Chen, Jianbin and Lin, Dennis KJ}]{winker2020construction}
Winker P, Chen J, Lin DK.
\newblock The construction of optimal design for order-of-addition experiment via threshold accepting.
\newblock In: Contemporary Experimental Design, Multivariate Analysis and Data Mining: Festschrift in Honour of Professor Kai-Tai Fang Cham: Springer; 2020.p. 93--109.

\bibitem[{Dueck and Scheuer(1990)Dueck, Gunter and Scheuer, Tobias}]{dueck1990threshold}
Dueck G, Scheuer T.
\newblock Threshold accepting: A general purpose optimization algorithm appearing superior to simulated annealing.
\newblock Journal of computational physics 1990;90(1):161--175.

\bibitem[{Claringbold(1955)Claringbold, PJ}]{claringbold1955use}
Claringbold P.
\newblock Use of the simplex design in the study of joint action of related hormones.
\newblock Biometrics 1955;11(2):174--185.

\bibitem[{Lewis et~al.(1991)Lewis, G. A. and Mathieu, D. and Phan-Tan-Luu, R.}]{lewis1991pharmaceutical}
Lewis GA, Mathieu D, Phan-Tan-Luu R.
\newblock Pharmaceutical Experimental Design.
\newblock 1st ed. Boca Raton, Florida, USA: CRC Press; 1991.

\bibitem[{Huisman et~al.(1984)Huisman, R and Van Kamp, HV and Weyland, JW and Doornbos, DA and Bolhuis, GK and Lerk, CF}]{huisman1984development}
Huisman R, Van~Kamp H, Weyland J, Doornbos D, Bolhuis G, Lerk C.
\newblock Development and optimization of pharmaceutical formulations using a simplex lattice design.
\newblock Pharmaceutisch Weekblad 1984;6:185--194.

\end{thebibliography}



\end{document}